# Vectorial adaptive optics for advanced imaging systems

Yifei Ma[†], Zimo Zhao[†], Jiahe Cui, Jingyu Wang, and Chao He[*]

*Department of Engineering Science, University of Oxford, Parks Road, Oxford, OX1 3PJ, UK*

[†]*These authors contributed equally to this work*
[*]*Corresponding author:* chao.he@eng.ox.ac.uk

**Vectorial adaptive optics (V-AO) is a cutting-edge technique extending conventional AO into the vectorial domain encompassing both polarization and phase feedback correction for optical systems. However, previous V-AO approaches focus on point correction. In this letter, we extend this AO approach into the imaging domain. We show how V-AO can benefit an aberrated imaging system to enhance not only scalar imaging but also the quality of vectorial information. Two important criteria, vectorial precision and uniformity are put forward and used in practice to evaluate the performance of the correction. These experimental validations pave the way for real-world imaging for V-AO technology and its applications.**

Adaptive optics (AO) is a powerful technology that has been used to correct wavefront distortions (introduced by phase aberrations) in optical systems [1–3]. It is applicable to various modern optical systems, spanning from microscopes to telescopes. However, besides traditional phase aberration, there exists other aberrations (see Fig, 1a) that would strongly affect the system performance such as polarization aberration (PA) [4–6]. PA can be introduced by various factors, including oblique beam incidence on birefringent optical elements, Fresnel's effect, as well as interactions with complex biological specimens or materials [7,8]. PA would imply distortions to the state of polarization (SOP), affecting the quality of polarimetric measurements. Complex polarimetry such as advanced Stokes vector/Mueller matrix microscopes therefore require vectorial adaptive optics (V-AO) to achieve the best performance, which can be implemented by manipulating the beam in the illumination and/or detection arms [9–12].

There are different ways to gain correction feedback for V-AO [4]. Here we use pupil intensity as reference [4] when correcting for an aberrated imaging system. Figure. 1b depicts a V-AO correction system where the V-AO module consists of spatial light modulators (SLMs) to achieve pixelated control of the SOPs and a deformable mirror (DM) to compensate for phase aberrations. Pupil intensity is provided as feedback to the V-AO module such that the correct pattern is used for compensation. In this work, we extend the cutting-edge V-AO technique from point correction to wide-field imaging. We first conduct PA correction for the whole field of view via V-AO and introduce two important criteria - vectorial information precision and vectorial uniformity - to evaluate the correction ability. A high-quality corrected vectorial field is then achieved and analysed. We perform V-AO correction on three different samples associated with PA – 1) air; 2) a polarization calibration sample; and 3) a birefringent crystal sample. A set of results before and after V-AO correction is presented for each sample to demonstrate the effectiveness of our V-AO method in correcting for aberrations in wide-field imaging systems.

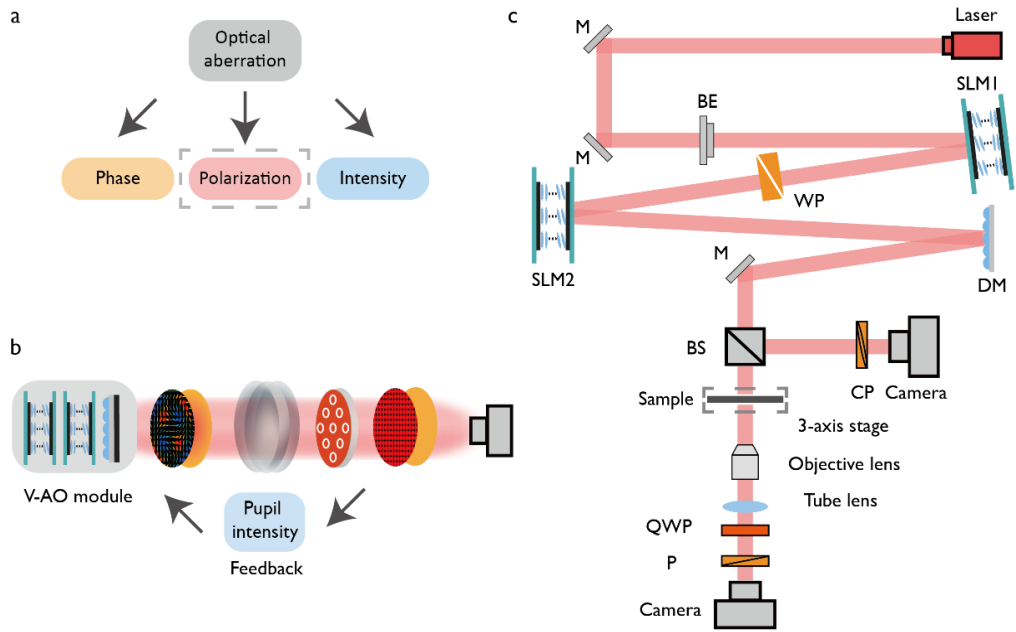

**Figure 1: Optical aberrations, V-AO correction strategy, and schematic of the experimental system.** (a) Different types of aberrations in the optical system: phase aberration, intensity aberration, and polarization aberration. (b) The V-AO correction strategy used in this work: a fixed polarizer filters the pupil intensity and provides feedback to the V-AO module for aberration compensation. (c) A schematic of the setup. M: mirror, BE: beam expander, SLM: spatial light modulator, WP: half-wave plate, QWP: quarter-wave plate, DM: deformable mirror, BS: beam splitter, CP: circular polarizer, P: linear polarizer. The majority of 4F systems are omitted for clarity.

**Setup and Correction Strategy**

To demonstrate the V-AO correction strategy in the imaging domain, we implement it in a wide-field Stokes imaging microscope. The schematic in Fig. 1 (c) illustrates the optical setup utilized for experimental validation. A HeNe laser (Melles Griot, 05-LHP-171, 632.8 nm) is used as the light source. After beam expansion, the illumination beam reflects off the first SLM (SLM1; Hamamatsu, X10468-01), passes through a half-wave plate (Thorlabs, WPH10M-633), and reflects off the second SLM (SLM2; Hamamatsu, X10468-01) before arriving at the DM (Boston Micromachines Corporation, Multi-3.5). The beam is then split by a 50:50 non-polarizing beam splitter (Thorlabs, BS010) into two paths. The reflection path passes through a circular polarizer which consists of a linear polarizer (Thorlabs, GL10-A) with the transmissive axis orientated in the horizontal direction and a quarter-wave plate (Thorlabs, WPQ10M-633) with the fast-axis oriented at 45 degrees. A camera (Thorlabs, DCC3240N) is placed after the equivalent circular polarizer to capture intensity images. The transmission path acts as the illumination source for the sample, the movement of which is controlled by a three-axis translation stage (Thorlabs, RBL13D). To realize Stokes vector measurements, a polarization state analyzer (PSA) is adopted, which consists of a quarter-waveplate (Thorlabs, WPQ10M-633) mounted into a motorized rotation stage (Thorlabs, PRM1/MZ8) and a linear polarizer (Thorlabs, GL10-A) with its transmissive axis oriented in the horizontal direction. The final images are recorded by another camera (Thorlabs, DCC3240N).

The V-AO SOP corrector consists of two SLMs (the first oriented at 0 degrees and the second at 45 degrees), to enable pixelwise generation of arbitrary spatially varying SOPs. Traditionally, each pixel of the SLM should first go through a retardance calibration process to determine the precise relation between retardance value and applied voltage for optimal control of the output SOP [10]. Here, the SOP correction strategy is achieved by circumventing this process, hence it allows us to conduct a correction procedure without calibration of the SLM. At the heart of our correction method is to find the best pixel values $p_1$, $p_2$ for every point in the pupil, where $p_1$, $p_2$ represent values applied to each of the two SLMs, respectively. Consider an arbitrary point at the pupil plane of the analysing channel (note in principle, any type of analysing component can be inserted in this channel with careful design to guide the V-AO corrector in displaying the optimal pre-compensation SOP), when the pixel values $p_1$, $p_2$ change on the SLMs, the resulting intensity will change as a function of $p_1$, $p_2$ and reach the highest value once the generated SOP matches the targeted SOP determined by the circular polarizer before the camera. Now the problem of correcting the SOP has been converted from finding the correct SLM retardance values to directly finding the best pixel values $p_1$, $p_2$ on the SLMs which result in the maximum intensity after the analyzer placed at the pupil plane.

The following procedures are adopted to find the best pixel values $p_1$, $p_2$ for every point in the pupil. The initial information required to start the correction is the pixel value boundary for each SLM, $p_1 \in [l_1, h_1]$ and $p_2 \in [l_2, h_2]$, where $p_i$ represents the value to be applied to each pixel on the SLM, and $l_i$ and $h_i$ are the lower and upper boundaries for each SLM. No explicit relationship between the pixel value and SLM retardance is required at this stage. The procedure starts with the derivation of a group of coarse scan values that take all possible ordered pairs from two fixed length integer arrays, the first extracted from the pixel value boundary of SLM1 and the second from that of SLM2, which is also the Cartesian product of these two arrays $[l_1, a, b, \cdots, h_1] \times [l_2, d, e, \cdots, h_2]$. Then, for each pair of coarse scan values $(p_1, p_2)$, we apply $p_1$ to all the pixels of SLM1 and $p_2$ to all the pixels of SLM2. An image of the pupil intensity profile is then captured. The above steps are then repeated until an intensity image is captured for all coarse scan value pairs. Next, for each point on the pupil, we plot the corresponding pupil intensity against $p_1$ and $p_2$. By interpolating for remaining pixel values within the boundaries, the optimal values $p_1$, $p_2$ for each pupil pixel can then be derived by locating the maximum value in the interpolation result. In our experiments, the pixel arrays used for the coarse scan process were $p_1 \in [8, 28, 48, \cdots, 128, 148]$ and $p_2 \in [44, 60, 76, \cdots, 140, 156]$. Both arrays take 8 steps to maintain a uniform interpolation region. It should be noted here that the SLM patterns calculated from the captured image may not be directly applicable to the two SLMs, as the number of pixels on the SLM may not be identical to that covered by the light beam on the camera. However, this can be easily overcome by performing another interpolation to map the calculated pattern back to the size of the SLM.

**Results**

System aberration correction is first performed using V-AO before imaging. We introduce two important criteria, vectorial precision ($P$) and vectorial uniformity ($U$), to evaluate the performance of the correction procedure. The SOP correction precision ($P$) is defined as the norm of the SOP vector difference on the Poincare sphere, and formulated as:

$$P = c\|S - \tilde{S}\| \quad (1)$$

where the target SOP is defined as $S = [S_1, S_2, S_3]$ and actual SOP defined as $\tilde{S} = [\tilde{S}_1, \tilde{S}_2, \tilde{S}_3]$, both of them are normalized and all parameters are within the range of $[-1, 1]$, as they represent vectors on the Poincare sphere. $c$ is a constant normalization factor for precision. To ensure that the correction precision ($P$) falls within the range of $[0, 1]$, a normalization factor of $c = \frac{1}{2}$ is used, as the norm of the SOP vector difference $\|S - \tilde{S}\|$ is in the range of $[0, 2]$. Note that this evaluation metric is valid over a line, a region of interest, or the whole pupil area.

To evaluate the correction ability further in the imaging domain, the uniformity of the resorted field is also examined and the SOP uniformity ($U$) is proposed as a new evaluation metric. $U$ is defined as the standard deviation of the actual SOP ($\tilde{S}$) from the mean SOP ($\bar{S}$) across the whole corrected area. The equation is expressed as:

$$U = c\sqrt{\overline{\|\tilde{S} - \bar{S}\|^2}} \quad (2)$$

where $c$ is a constant normalization factor for uniformity. To ensure that the correction precision ($U$) falls within the range of $[0, 1]$, a normalization factor of $c = 1$ is used due to the symmetrical nature of Stokes parameter values. Each parameter $\bar{S} = [\bar{S}_1, \bar{S}_2, \bar{S}_3]$ is defined as the mean of parameters $\tilde{S} = [\tilde{S}_1, \tilde{S}_2, \tilde{S}_3]$ across the whole pupil, as expressed in equation (3) below. Note that this evaluation metric is valid over the whole imaging pupil or sub-regions. Hence, this metric can be used to evaluate global uniformity.

$$\bar{S}_i = \overline{\tilde{S}_i} \quad (3)$$

After introducing the two new evaluation metrics, we conducted the systematic aberration correction procedure via V-AO. Fig. 2(a) illustrates the Stokes vector field of the pupil along with their Stokes parameters before and after the correction process. From the sub-figures, it can be observed that the SOPs in the original pupil area are both incorrect and non-uniform compared to the target SOP, while our V-AO approach can provide an excellent correction resulting in a uniform targeted SOP profile across the whole pupil area. We then conduct quantitative comparison by selecting and analysing SOPs with a random cross-section (see Figs. 2(a) and 2(b)). It can be found that, despite the significant distortion in $S_3$, our method can still perform high quality Stokes vector corrections. We then drew upon parameters $P$ and $U$ to further evaluate the correction ability. Their values before and after correction are calculated and illustrated in Fig. 2(c). We can see that the vectorial precision has been greatly enhanced from an initial <1% to >95% as depicted by the pink and red lines, indicating the significance of our V-AO method in correcting vectorial information. Meanwhile, the uniformity of the Stokes vector field is also boosted from 77.1% to 91.2%. These results validate that our V-AO correction process can achieve good performance in correcting for spatially varying PAs in the imaging domain.

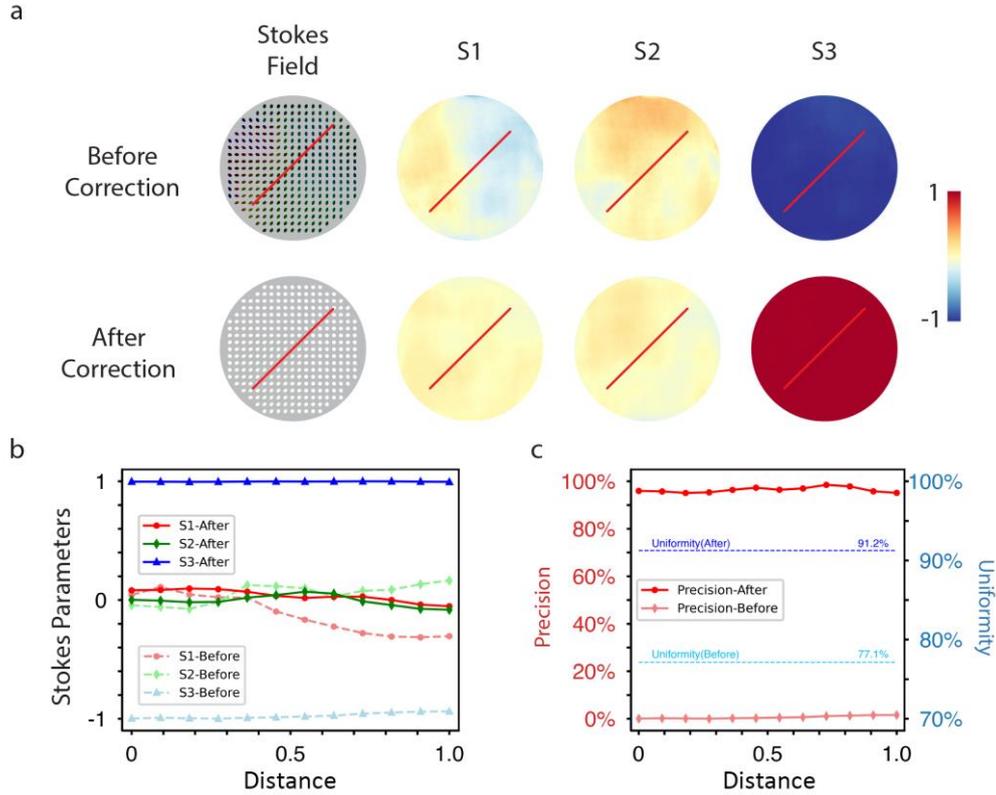

**Figure 2: Stokes vector field analysis before and after V-AO correction.** (a) Stokes vector fields and individual Stokes parameters before and after V-AO correction. (b) The variation of each Stokes parameter before and after correction along the red line in Fig. 2(a). (c) The values of $P$ (vectorial precision) and $U$ (vectorial uniformity) were calculated along the same line before and after correction.

We then demonstrate the effectiveness of our V-AO method by applying it to important real-world scenarios. The experiments were performed in a) air; b) a polarisation calibration sample; and c) a birefringent crystal sample – as shown in Fig. 3. In Figs. 3(a), 3(b), and 3(c), the columns represent the sample schematic (column 1), and the Stokes vector (large) and intensity (small) fields of the target SOP under the following conditions: ground truth (column 2), aberrations present with V-AO off (column 3), and aberrations present with V-AO on (column 4). In the case of the air sample, we can infer from the ground truth to expect uniform circular SOP and a bright intensity image. Comparison between V-AO off and on conditions show enhancement of the vectorial information precision, uniformity, and pupil intensity. It should be noted that 1) the ground truth is taken from a traditional low-resolution Stokes vector microscope, where system induced aberrations such as Fresnel's effects which can originate from high numerical aperture lenses [13,14] can be omitted; 2) the original intensity is not perfectly uniform, which happens in traditional wide-field imaging. It can be compensated via beam shaping or further adaptive correction techniques; 3) the intensity images are captured via circular SOP illumination and right-hand elliptical SOP analysis only (QWP oriented at 128.3 degrees and P oriented at 0 degrees), as an example. In the latter two samples, we used a standard QWP' as the calibration target, and a thin film of uniaxial positive crystal sample featuring different retardance and fast axis orientation distributions as the birefringent sample. Comparison with the ground truth validates the effectiveness of the correction process. Notably, in the case of the birefringent sample, it can be more appreciated how structural information can be revealed by the corrected vectorial and intensity images, further demonstrating the capability of our V-AO correction technique.

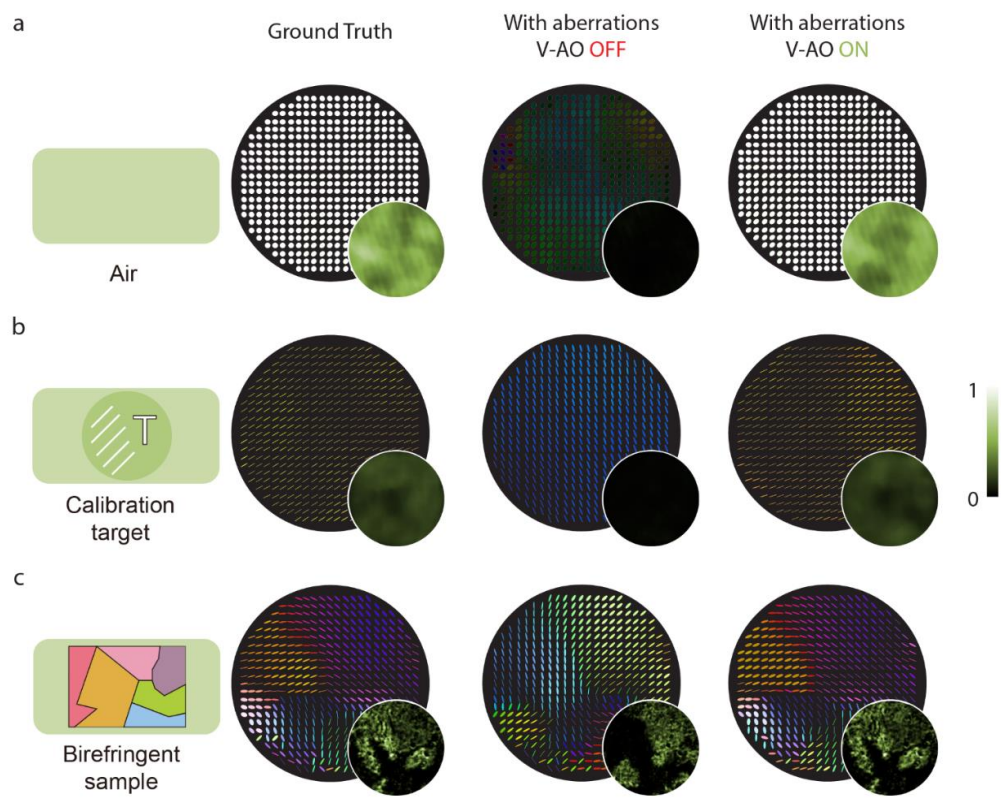

**Figure 3: V-AO correction for aberrated imaging systems.** Stokes vector field images and intensity images of three different samples: (a) air, (b) a calibration target sample, and (c) a birefringent sample. The following scenarios are demonstrated: ground truth, aberration present with V-AO off, and aberration present with V-AO on. In (a), the sample is air, hence the SOP and intensity should be the same as the input. We can see that the uniform circular SOP and intensity level can be corrected from V-AO off to V-AO on. In (b), as the calibration target is a QWP, the obtained Stokes vector field changes to linear SOP. The intensity also changes using the same polarization state analyzer. In (c), a complex birefringent sample is used. With V-AO on, the spatially varying SOP and intensity agree well with the ground truth.

**Discussion**

In this letter, we extended V-AO from point correction to the imaging domain, through the validation of a widely used polarization-sensitive Stokes vector microscope system. Two important criteria have been put forward – vectorial precision $P$ and uniformity $U$ – to evaluate V-AO correction ability. They have been used to quantitatively analyse the performance of our V-AO correction method in a wide-field imaging system, demonstrating a precision of >95% and uniformity of >91%. We then use the V-AO system which features complex geometry with intrinsic vectorial aberrations to perform real-world imaging correction. We used three typical samples to conduct the experiments including air, a polarisation calibration target, and a birefringent sample. The first one was used to validate the system aberration correction performance of the original illumination SOP (right-hand circular SOP) compared with the ground truth, while the latter two are used to demonstrate target SOP recovery for uniform and spatially varying SOPs, respectively. Both the vectorial images and intensity images validated that the V-AO performed well in the imaging domain, and it can recover both fine vectorial polarisation structures and sample intensity features correctly.

The domain shift from a point to a field paves the way for various AO-targeted imaging scenarios including super-resolution imaging and vectorial imaging. For wide-field super-resolution imaging methods like SIM [15] and SDOM [16], we can now not only correct for both phase and polarization perturbations in the structured illumination light, but also competently design new V-AO strategies for the detection path. For traditional vectorial imaging methods such as Stokes vector/Mueller matrix imaging suffering from spatially varying aberrations [5], e.g., those induced by high numerical aperture objective lenses or the specimens themselves, V-AO has now been taken a step further to enable novel V-AO enhanced vectorial imaging methodologies, with the expectation to benefit various fields such as biomedical and clinical characterization and galaxy detection.

Furthermore, there are potential research directions on the software side. 1) The efficiency and precision of the V-AO correction algorithm can again be improved with respect to the quantitative criteria $P$ and $U$. Currently, the algorithm is based upon finding the highest intensity value, but there are intriguing scopes that we can investigate further, such as combining cutting-edge machine learning techniques with heuristic algorithms [17] to improve the speed and accuracy of correction under the guidance of metrics $P$ and $U$. 2) The V-AO correction method we adopted here is a type of quasi-sensorless method, which is currently an approach that does not involve the concept of modes [2] – while it is highly likely that vectorial modes [18] will be developed and harnessed in the future to enhance our correction performance. In parallel, on the hardware side there are also many avenues we can explore. 1) Currently the CP positioned at the conjugate pupil plane is fixed and uniform. While it is already very useful for ensuring uniform SOP illumination, in the future the CP can be replaced by pixelated tuneable devices such as cascaded LC devices [19], or LC-based metamaterials [20]. Then, the input light beam can be manipulated to exhibit more complex spatial patterns, potentially useful for advanced sample probing. 2) With an increased number of AO devices, the V-AO format can be further extended for use in imaging scenarios. The current system using two SLMs can provide pre-aberration compensation for a fixed SOP illumination beam; furthermore, in the detection arm, we can also harness four SLMs to permit the transformation between arbitrary-to-arbitrary polarization states and phase profiles, enabling AO correction for the detection optical pathway [11]. 3) The current V-AO strategy controls polarization only through time manipulation. However, there is good potential to increase the degrees of freedom [21] for polarization control using additional AO correctors. Perhaps another interesting observation here is the intensity variation before and after correction. Apart from the intensity non-uniformity induced by the light source itself, it may be induced by the V-AO corrector. Although this does not affect the uniform values of the Stokes vector field during imaging, if absolute intensity needs to be considered in the future, beam shaping or more degrees of freedom controlled by AO would be a promising direction to explore further.

Overall, V-AO-enhanced vectorial imaging has for the first time been explored and analysed. Future benefits are expected in a wide range of applications, spanning from astronomical telescopes to microscopy.